\documentclass[11pt,a4paper]{article}
\usepackage[margin=27mm]{geometry}
\usepackage[UTF8,fontset=fandol,scheme=plain]{ctex}
\usepackage{amsmath,amssymb,amsthm,booktabs,graphicx,array}
\usepackage[hidelinks]{hyperref}
\AtBeginDocument{}
\newtheorem{proposition}{Proposition}
\newcommand{\logit}{\operatorname{logit}}
\title{China's Local Government Debt, Strategic Competition, and Bitcoin Demand:\\ A Conditional Transmission Framework}
\author{Bit Sun\\Independent Researcher}
\date{November 11, 2026}
\begin{document}
\maketitle
\begin{abstract}
How might the resolution of China's local-government debt affect strategic competition and Bitcoin demand? This paper develops a conditional framework connecting debt accounting, bank balance-sheet adjustment, public spending priorities, and constrained portfolio substitution. A debt swap can increase recognized government debt without increasing consolidated principal, while its effect on credit capacity depends on capital relief, loss recognition, and recapitalization. A protected strategic-spending floor can preserve current effort while shifting adjustment toward civilian productive services. In the asset block, stronger relative preference for Bitcoin does not ensure greater effective demand: access constraints and global funding conditions can reverse the sign. We derive a finite-shock threshold and deterministic sign bounds, and distinguish Bitcoin demand from demand for dollar stablecoins. Reproducible numerical illustrations and channel-removal exercises show that opposite outcomes are compatible with the same assumed fiscal stress. The analysis uses dated primary sources to establish institutional context, but estimates neither a debt-to-Bitcoin causal coefficient nor a price-impact scale. Its contribution is a compact set of conditions and measurement distinctions for evaluating transmission claims, rather than an unconditional price forecast or an inference about conflict probabilities.
\end{abstract}
\medskip

\section{Introduction}

China's local-government debt raises a question that cannot be answered by the size of a balance sheet alone: who absorbs the adjustment, and through which decisions does it reach international competition and financial markets? A refinancing operation, a bank recapitalization, a reduction in public services, and a household portfolio response have different resource and valuation consequences. Treating them as interchangeable steps obscures the conditions that determine the final outcome.

RAND's recent study connects local-debt measurement, institutional responses, contrasting city cases, and adjustment scenarios to U.S.-China competition. It does not equate local fiscal strain with an immediate reduction in national strategic effort \cite{rand}. The present paper extends that motivating inquiry to a conditional Bitcoin mechanism; the report does not identify this transmission.

Three distinctions organize the argument. First, the recognition of a liability differs from the creation of a new liability. An exchange of local-government financing vehicle (LGFV) obligations for government bonds can change the recorded stock and creditors' risk weights without canceling aggregate principal. Second, current strategic spending differs from the economic base that sustains future competition. An allocation rule can protect a spending priority while imposing costs on public services and private production. Third, demand for assets outside a domestic currency differs from effective Bitcoin demand. Investors can choose dollar claims, lose access to trading services, or liquidate risky positions at the same time that their concern about domestic fiscal conditions increases.

The financial literature provides reasons to keep these distinctions explicit. Huang, Pagano, and Panizza find historical crowding out of private investment through local funding constraints \cite{huang}. Li and coauthors, studying the 2015 debt-to-bond swap, identify a countervailing risk-weighting channel that supports private credit and productivity \cite{li}. These studies concern different interventions and samples; they should not be turned into an invariant coefficient on all public debt. On the asset side, Che and coauthors document the importance of U.S. monetary policy and institutional risk taking for crypto markets \cite{che}. Recent work by Drexler, Guettler, and Sun also finds increasing broad-equity exposure for Bitcoin \cite{drexler}. Neither result identifies the effect of Chinese local debt.

Existing research already examines Chinese policy uncertainty and Bitcoin. Cai and coauthors report time- and frequency-dependent relationships rather than a universal hedge property \cite{cai}. The present contribution is consequently narrower than a discovery that Chinese uncertainty matters for cryptoassets. It joins liability recognition and loss allocation to protected spending and a constrained BTC--stablecoin allocation block. The resulting propositions state what follows from the assumptions and what remains unidentified. The elementary accounting and optimization results are not presented as new general theorems.

The paper has two completed analytical components. The first shows how a debt resolution can preserve principal yet change bank credit headroom and the composition of public spending. The second derives effective-demand and price-pressure sign conditions with separate access, preference, and global-funding channels. The components share a stress variable, but the policy responses are conditional primitives: the framework is not a jointly estimated fiscal-monetary equilibrium. Numerical exercises illustrate its restrictions and counterexamples. No probability is attached to a scenario, and no realized Bitcoin return is attributed to a local-debt announcement.

\section{Debt measurement and institutional context}
\label{sec:measurement}

The empirical starting point must retain the date, perimeter, and reporting vintage of each liability measure. Table~\ref{tab:perimeter} records four primary-source observations. The Ministry of Finance (MOF) measure concerns recognized local-government liabilities. The RAND measure additionally covers identifiable LGFV interest-bearing obligations. The IMF's later departmental paper uses a broader local-government and LGFV concept for a different year \cite{mof2024,mof2026,rand,imf2026}. The rows are not rival estimates of one identically defined random variable, and their spread is not a statistical confidence interval.

\begin{table}[htbp]
\centering
\small
\caption{Dated liability measures with different perimeters}
\label{tab:perimeter}
\begin{tabular}{>{\raggedright\arraybackslash}p{0.20\linewidth}>{\raggedright\arraybackslash}p{0.18\linewidth}>{\raggedright\arraybackslash}p{0.22\linewidth}>{\raggedright\arraybackslash}p{0.29\linewidth}}
\toprule
Source and vintage & Observation date & Reported magnitude & Coverage and interpretation\\
\midrule
MOF final accounts, July 2025 & End-2024 & RMB 47.540125 trillion & Recognized local-government debt; all LGFV liabilities are not included.\\[4pt]
RAND, October 2026 & 2024 & About USD 12 trillion; 64\% of GDP & Official local debt plus identifiable LGFV interest-bearing debt; source estimate, not replicated here.\\[4pt]
IMF DP 2026/019 & End-2025 & About 90\% of GDP & Official local-government and LGFV debt under the study's broader coverage.\\[4pt]
MOF monthly release, October 2026 & End-September 2026 & RMB 60.6539 trillion & Recognized local-government debt; not a fresh estimate of total local plus LGFV liabilities.\\
\bottomrule
\end{tabular}
\end{table}

The MOF's September 2026 release also reports an average remaining bond maturity of 10.9 years and an average outstanding bond interest rate of 2.70 percent \cite{mof2026}. These figures describe government bonds, not the entire cost and maturity profile of LGFVs. An aggregate interest rate cannot recover the cash flows of distressed entities. Nor can an increase in recognized debt between 2024 and 2026 be interpreted entirely as new real expenditure: refinancing, liability recognition, and fresh borrowing must be separated.

The November 2024 program illustrates why that separation matters. The authorities approved RMB 6 trillion of additional quota over 2024--2026 to replace existing hidden debt and described a further RMB 4 trillion allocation from special-bond quotas over five years \cite{swap}. A quota is neither a realized disbursement nor a transfer to households. The conversion of an existing obligation is also distinct from monetary issuance. The relevant questions are which contracts change, which entities remain liable, and whether the resources released by lower debt service finance new expenditure or repair balance sheets.

The World Bank emphasizes the role of weak land revenue and regional differences in fiscal adjustment \cite{wb}. The IMF's 2026 study further stresses interconnected local entities, bank preparation, insolvency tools, and central coordination \cite{imf2026}. These mechanisms motivate attention to creditor losses and local productive capacity. They do not establish a specific aggregate loss or an inevitable banking crisis.

Asset access is another institutional constraint. The February 2026 mainland notice on virtual-currency risks restricts relevant business and payment services and replaces the 2021 notice \cite{notice}. Mainland and Hong Kong arrangements should therefore not be collapsed into a frictionless national investment market. The model below neither assumes universal domestic access to Bitcoin nor treats on-chain activity as evidence of a user's nationality.

\section{Debt resolution, bank capacity, and strategic resources}

\subsection{A fixed-perimeter accounting identity}

Fix a perimeter consisting of local governments and covered LGFVs. Eliminate claims between entities inside that perimeter. Let $E$ denote recognized debt owed outside it and $V$ covered LGFV debt owed outside it that has not already been included in $E$. The variables are nonoverlapping principal stocks; they are not a direct transcription of every published gross-debt series. A one-for-one replacement of amount $x\in[0,V]$ yields
\begin{equation}
 E^{+}=E+x,\qquad V^{+}=V-x.
 \label{eq:swap}
\end{equation}
This construction holds fresh borrowing, write-offs, cash repayment, and changes in the covered population at zero. Under those restrictions, it identifies a reclassification and contractual exchange rather than a decline in total principal.

\begin{proposition}[Recognition and conditional credit headroom]
\label{prop:swap}
Under \eqref{eq:swap}, $E^{+}+V^{+}=E+V$. At unchanged principal, replacing an interest rate $r_V$ by $r_E$ reduces annual contractual interest by $x(r_V-r_E)$. For a bank facing a binding capital constraint with required ratio $c_b>0$, private-loan weight $w_P>0$, and exposure weights $w_V,w_E$, a swap of bank exposure $x$, net capital burden $H$, and recapitalization $J$ changes private-credit headroom by
\begin{equation}
 \Delta P_{\rm cap}
 =\frac{J-H+c_b(w_V-w_E)x}{c_b w_P}.
 \label{eq:bank}
\end{equation}
It expands headroom exactly when $J+c_b(w_V-w_E)x>H$.
\end{proposition}
\begin{proof}
Adding \eqref{eq:swap} cancels $x$. The interest statement follows by subtracting the old and new contractual payments on the replaced principal. Initially write the bank constraint as $K=c_b(w_V V_b+w_P P+A)$, where $A$ denotes other risk-weighted exposures. Let $L$ be the regulatory-capital loss, $\Delta A$ the associated change in other risk-weighted exposures, and define $H=L+c_b\Delta A$. Recapitalization proceeds are assumed to enter as zero-risk cash before lending. After adjustment the constraint is $K+J-L=c_b[w_V(V_b-x)+w_E x+w_P(P+\Delta P_{\rm cap})+A+\Delta A]$. Subtraction gives \eqref{eq:bank}; division is valid because $c_b w_P>0$.
\end{proof}

Equation~\eqref{eq:bank} describes capacity conditional on sufficient funding and private loan demand. It is not an unconditional lending response. The net burden $H$ includes the offset from changes in capital requirements and need not equal the gross loss $L$. Loss recognition can dominate regulatory relief; recapitalization can restore capacity; and a bank that is not capital constrained need not expand lending. A reduction in contractual interest income is not automatically a capital loss of the same size: expected default losses, funding costs, discounting, and duration also matter. This distinction accommodates both the historical crowding-out and debt-composition evidence without treating them as contradictory universal claims \cite{huang,li}.

For example, take entirely hypothetical normalized values $x=1$, $c_b=0.1$, $w_V=0.5$, $w_E=0.2$, and $w_P=1$. Capital relief is $0.03$. With $H=0.01$ and $J=0$, headroom increases by $0.20$. With $H=0.08$ it decreases by $0.50$; adding $J=0.10$ instead makes the change $0.50$. These numbers demonstrate the threshold, not the risk weights, losses, or credit response of the current Chinese program.

The accounting perimeter remains essential. If public banks are brought into a wider consolidation, claims between public creditors and debtors must also be eliminated, while deposits and liabilities to outside creditors remain. Neither $E+V$ nor the swap saving is automatically national net debt or a net social resource gain. Principal conservation also does not establish solvency: the market value of underlying projects and future fiscal capacity may have changed.

\subsection{A protected allocation rule}

Let $z\geq0$ denote fiscal stress under a specified policy regime. It is a conceptual state variable, not a debt-to-GDP ratio with an estimated numerical scale. Suppose the resources available after fiscal absorption are
\begin{equation}
 F(z)=B-c_z z,\qquad c_z\geq0.
 \label{eq:envelope}
\end{equation}
The slope $c_z$ is a policy-dependent net resource cost, incorporating transfers, restructuring, and other adjustments; it is not the stock of local debt. A central budget, bank balance sheet, and local budget are distinct, so \eqref{eq:envelope} is a conditional reduced-form envelope rather than an assertion that every local liability is centrally funded.

Allocate strategic effort $S$ and civilian productive services $G$ by
\begin{equation}
 \max_{S,G}\ \alpha\log S+(1-\alpha)\log G
 \quad\text{s.t.}\quad S+G=F,\ S\geq\underline S,\ G\geq\underline G,
 \label{eq:allocation}
\end{equation}
where $0<\alpha<1$ and both floors are positive. The problem is feasible only for $F\geq\underline S+\underline G$. A strategic floor represents an imposed short-run priority; it is not a measured Chinese spending guarantee.

\begin{proposition}[Protection changes the incidence of adjustment]
\label{prop:budget}
For feasible $F$, the unique allocation is
\begin{equation}
 S^*(F)=\min\{\max\{\alpha F,\underline S\},F-\underline G\},
 \qquad G^*(F)=F-S^*(F).
 \label{eq:clamp}
\end{equation}
Away from transition points and the feasibility boundary, $dS^*/dz$ equals $-\alpha c_z$ in the interior, zero at a binding strategic floor, and $-c_z$ at a binding civilian floor. At the strategic floor, $dG^*/dz=-c_z$.
\end{proposition}
\begin{proof}
Substitute $G=F-S$. The objective is strictly concave on the feasible interval $[\underline S,F-\underline G]$. Its unconstrained first-order condition is $\alpha/S=(1-\alpha)/(F-S)$, giving $S=\alpha F$. Projection onto that interval yields \eqref{eq:clamp}. Differentiating each active branch using \eqref{eq:envelope} proves the slopes. At the strategic floor $S$ is constant, so $G=F-\underline S$ absorbs the full marginal change.
\end{proof}

\begin{figure}[htbp]
\centering
\includegraphics[width=0.92\linewidth]{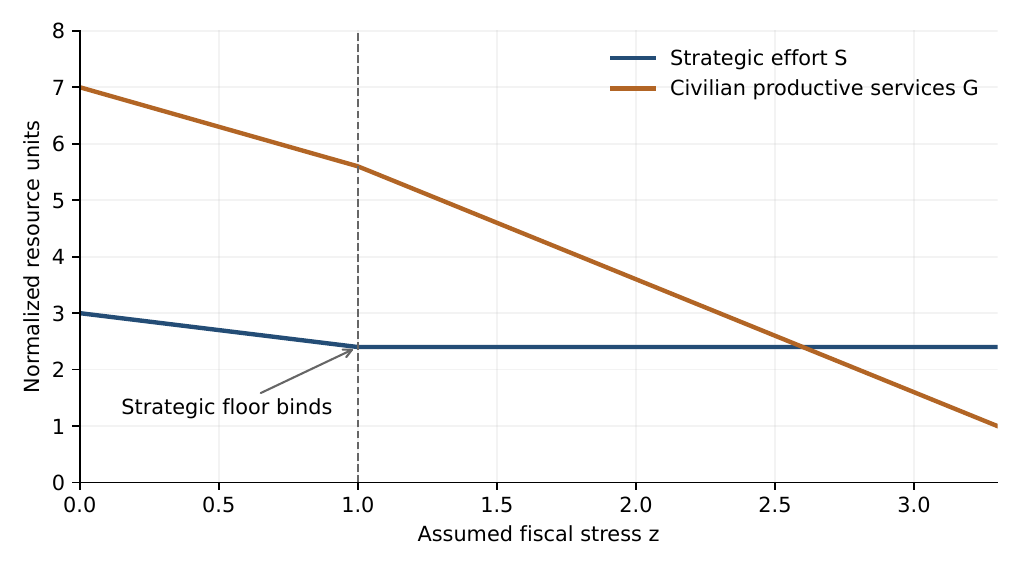}
\caption{Hypothetical allocation under a protected strategic floor. Parameters are $B=10$, $c_z=2$, $\alpha=0.3$, $\underline S=2.4$, and $\underline G=1$. The strategic floor binds at $z=1$; the illustrated domain ends at the feasibility boundary $z=3.3$. Units are normalized resources, not currency amounts or measured defense capacity.}
\label{fig:budget}
\end{figure}

Figure~\ref{fig:budget} shows the assumed allocation, not observed budgets. Strategic spending initially falls with the envelope, then remains at its floor while civilian productive services continue to decline. More generally, the branch depends on the floor levels; a civilian floor can instead shift the marginal cut toward strategic effort. The example does not assert that one branch must describe China.

To separate current effort from its productive base, suppose only for interpretation that next-period capacity satisfies $K_{t+1}=(1-\delta)K_t+\theta G_t-\chi z_t$, with $\theta,\chi\geq0$. Protecting $S_t$ can then coexist with weaker future capacity. This additional relation is assumed, not estimated. Relative U.S.-China capability would also depend on U.S. choices, technology, productivity, and the conversion of spending into effective capability. The allocation result therefore supports no inference about a conflict probability or a country's willingness to initiate conflict.

\section{Portfolio access, stablecoins, and Bitcoin pressure}
\label{sec:assets}

\subsection{Reference-price demand shifters}

Consider a marginal group of potential allocators affected by, or responding to news about, the fiscal state. Let $W(z)=W_0 e^{\omega z}$ be their intended dollar-value allocation to a stylized BTC--dollar-stablecoin basket at reference prices. The elasticity $\omega$ can be positive or negative: concern about domestic claims can increase desired reallocation, whereas wealth losses can reduce it. The group is not identified with all Chinese savers, and $W$ is not measured capital flight.

Only a fraction $a(z)=a_0 e^{-\kappa z}$ is accessible, where $0<a_0\leq1$ and $\kappa\geq0$. Inaccessible or unchosen funds remain in an outside option. Within the accessible basket, set
\begin{equation}
 p(z)=\frac{1}{1+\exp[-\logit(p_0)-\eta z]},\qquad
 \logit(p_0)=\log\frac{p_0}{1-p_0},\quad 0<p_0<1.
 \label{eq:share}
\end{equation}
Here $p$ is a reference-price BTC share. The logistic form is a parsimonious relative-choice restriction, consistent with a logit specification for a utility difference that changes by $\eta z$. It does not estimate that utility difference, and $\eta$ may have either sign. A preference for nonsovereign assets is only one possible response; a preference for dollar liquidity is another.

Define the reference-price demand shifters
\begin{equation}
 Q_B(z)=a(z)W(z)p(z),\qquad
 Q_D(z)=a(z)W(z)[1-p(z)].
 \label{eq:demand}
\end{equation}
They are dollar-value quantities at common reference valuations, not observed net orders, cross-border flows, or numbers of coins. Normalizing by the positive baseline gives
\begin{align}
 q_B(z)&=\frac{Q_B(z)}{Q_B(0)}
       =e^{(\omega-\kappa)z}\frac{p(z)}{p_0},\label{eq:qb}\\
 q_D(z)&=\frac{Q_D(z)}{Q_D(0)}
       =e^{(\omega-\kappa)z}\frac{1-p(z)}{1-p_0}.\label{eq:qd}
\end{align}
Each ratio uses its own baseline; their changes should not be added as if the denominators were identical. Their levels are also not market-wide capitalization measures. Segmented access and limits to arbitrage give reasons to distinguish a regional demand shifter from a global price \cite{makarov}.

\begin{proposition}[Relative preference and absolute demand can diverge]
\label{prop:demand}
For $a_0,W_0>0$,
\begin{align}
 \frac{d\log Q_B}{dz}&=\omega-\kappa+\eta[1-p(z)],\label{eq:derivb}\\
 \frac{d\log Q_D}{dz}&=\omega-\kappa-\eta p(z),\label{eq:derivd}\\
 \frac{d\log(Q_B/Q_D)}{dz}&=\eta.\label{eq:odds}
\end{align}
Thus $\eta>0$ does not ensure that either absolute demand increases. If $\eta>0$ and $0<\kappa-\omega<\eta$, $Q_B$ has at most one interior maximum on $z\geq0$. It occurs, when the solution is positive, at
\begin{equation}
 p^*=1-\frac{\kappa-\omega}{\eta},\qquad
 z^*=\frac{\logit(p^*)-\logit(p_0)}{\eta}.
 \label{eq:turn}
\end{equation}
\end{proposition}
\begin{proof}
Equation~\eqref{eq:share} gives $p'=\eta p(1-p)$. Log-differentiating \eqref{eq:demand} proves \eqref{eq:derivb}--\eqref{eq:odds}. For $\eta>0$, the BTC log-demand slope has derivative $-\eta^2 p(1-p)<0$. A zero slope therefore gives a unique maximum if it lies in the domain. Solving the zero-slope condition gives \eqref{eq:turn}. If the proposed $z^*$ is negative or the initial slope is nonpositive, no positive interior maximum is implied.
\end{proof}

The proposition separates a relative hedge narrative from an executable allocation. For example, if $\eta>0$ but $\kappa>\omega+\eta[1-p(z)]$, both demand shifters decrease even while the BTC-to-stablecoin ratio increases. Preference evidence by itself cannot establish net purchases. Conversely, when $\eta<0$, intended foreign-currency allocation can move primarily toward dollar claims. BIS research on dollar-denominated stablecoins provides an institutional reason to examine that alternative, rather than equating cryptoasset adoption with a decline in dollar use \cite{bis}.

\subsection{From demand shifts to a conditional price sign}

At a reference valuation, let $N(z)$ denote the net demand shift from the rest of the market, with $N(0)=0$. It includes favorable liquidity effects net of liquidation, funding, custody, and other demand effects not already counted in $Q_B$. A first-order market-clearing approximation is
\begin{equation}
 \Delta\log P_B\ \simeq\
 \frac{Q_B(z)-Q_B(0)+N(z)}{M},\qquad M>0.
 \label{eq:price}
\end{equation}
Here $M$ is the dollar-value response of aggregate net demand to a log-price change. It incorporates price elasticity and accessible supply; it is not mechanically equal to market capitalization or a displayed order-book depth. Equation~\eqref{eq:price} is a local approximation around reference prices, not a globally valid valuation formula. Funding constraints can change both demand and liquidity \cite{brunnermeier}, while macroeconomic exposures motivate allowing the rest-of-market term to vary \cite{che,drexler}.

For a transparent comparison, impose the illustrative restriction $N(z)=-Q_B(0)gz$, with $g$ a constant net global drag. Negative $g$ represents net support. Define
\begin{equation}
 I(z)=q_B(z)-1-gz.
 \label{eq:index}
\end{equation}
The index is dimensionless demand pressure in baseline-$Q_B$ units. It is not a predicted return. Under \eqref{eq:price}, its price scale would be the unestimated positive factor $Q_B(0)/M$; the exercise below reports no value for that factor.

\begin{proposition}[Finite-shock sign and dollar substitution]
\label{prop:threshold}
For fixed $z>0$, the index is positive exactly when
\begin{equation}
 g<g^*(z):=
 \frac{e^{(\omega-\kappa)z}p(z)/p_0-1}{z}.
 \label{eq:threshold}
\end{equation}
Its initial slope is $I'(0)=\omega-\kappa+\eta(1-p_0)-g$. A rising stablecoin demand shifter can coexist with negative marginal BTC pressure: sufficient and necessary conditions within this specification are
\begin{equation}
 \omega-\kappa-\eta p(z)>0,\qquad
 g>q_B(z)\{\omega-\kappa+\eta[1-p(z)]\}.
 \label{eq:coexist}
\end{equation}
\end{proposition}
\begin{proof}
Substitution of \eqref{eq:qb} into \eqref{eq:index}, followed by division by $z>0$, gives \eqref{eq:threshold}. Since $q_B(0)=1$, \eqref{eq:derivb} gives the initial slope. Stablecoin demand rises exactly when its log derivative \eqref{eq:derivd} is positive. Differentiating \eqref{eq:index} gives $I'=q_B\{\omega-\kappa+\eta(1-p)\}-g$, proving \eqref{eq:coexist}.
\end{proof}

The coexistence statement concerns a marginal change; it should not be substituted for the finite-shock condition \eqref{eq:threshold}. For a concrete possibility at $z=0$, let $\omega-\kappa=0.5$, $\eta=0.2$, $p_0=0.2$, and $g=0.8$. Stablecoin log-demand grows at $0.46$, while $I'(0)=-0.14$. No positive stablecoin dollar-price return follows: a maintained peg concerns price, whereas this result concerns desired balances. Issuance, redemption, exchange turnover, and reserve-asset holdings are also distinct measurements.

Access has an additional level effect that normalization can hide. If $a_0=0$, both raw local demand shifters in \eqref{eq:demand} are zero, and $q_B$, $q_D$, and $g$ normalized by $Q_B(0)$ are undefined. Equation~\eqref{eq:price} must then be used directly; global demand can still move the price. Taking $a_0$ toward zero with a fixed dollar-valued $N$ differs from holding the normalized $g$ fixed. This limiting case prevents a large normalized index from being mistaken for a large aggregate price impact.

\subsection{Deterministic bounds rather than probabilities}

\begin{proposition}[Conditional sign bounds]
\label{prop:bounds}
Fix $z>0$ and $p_0\in(0,1)$. On a rectangular assumed parameter set
\begin{equation}
 \omega\in[\omega_L,\omega_U],\quad
 \kappa\in[\kappa_L,\kappa_U],\quad
 \eta\in[\eta_L,\eta_U],\quad
 g\in[g_L,g_U],
 \label{eq:box}
\end{equation}
$I$ is increasing in $\omega$ and $\eta$ and decreasing in $\kappa$ and $g$. Its minimum is obtained at $(\omega_L,\kappa_U,\eta_L,g_U)$, and its maximum at $(\omega_U,\kappa_L,\eta_U,g_L)$. A positive sign is robust on the set only if $I_{\min}>0$; a negative sign is robust only if $I_{\max}<0$. If both signs occur, the set does not identify a sign.
\end{proposition}
\begin{proof}
Direct differentiation gives $I_\omega=zq_B>0$, $I_\kappa=-zq_B<0$, $I_g=-z<0$, and $I_\eta=zq_B(1-p)>0$. Continuity and coordinate monotonicity give the extrema at the stated corners. The sign statements follow from those bounds.
\end{proof}

These are deterministic sensitivity bounds on assumed inputs, not confidence intervals. Without defensible bounds on the parameters and the global term, fiscal stress alone supplies no sign identification. Even observing Bitcoin prices cannot separately identify all channels: in \eqref{eq:price}, different $Q_B$ paths can be offset by different $N$ paths. Estimating a probability of appreciation would additionally require an identified distribution of states and parameters, which the framework does not supply.

\section{Reproducible illustrations and channel removals}

Take $z=1$, $p_0=0.2$, $\eta=0.8$, and $\omega=0.4$, so that the imposed BTC share becomes $p(1)=0.357486$. Table~\ref{tab:scenarios} changes only access erosion and global drag. Scenario names summarize imposed conditions. They are neither historical episodes nor estimated forecasts of China's adjustment.

\begin{table}[htbp]
\centering
\small
\caption{Hypothetical finite-shock demand shifts and pressure}
\label{tab:scenarios}
\begin{tabular}{lrrrrr}
\toprule
Scenario & $\kappa$ & $g$ & $q_B-1$ & $q_D-1$ & $I$ \\
\midrule
Orderly adjustment & 0.2 & 0.1 & 1.183 & -0.019 & 1.083 \\
Global liquidation & 0.2 & 1.5 & 1.183 & -0.019 & -0.317 \\
Access tightening & 1.2 & 0.1 & -0.197 & -0.639 & -0.297 \\
Mixed adjustment & 0.6 & 0.2 & 0.463 & -0.342 & 0.263 \\
\bottomrule
\end{tabular}

\par\medskip\footnotesize
Each $q$ uses its own asset's baseline. $I=q_B-1-g$ has no estimated mapping to a Bitcoin return. All inputs are assumed; values are rounded to three decimals.
\end{table}

\begin{figure}[htbp]
\centering
\includegraphics[width=0.92\linewidth]{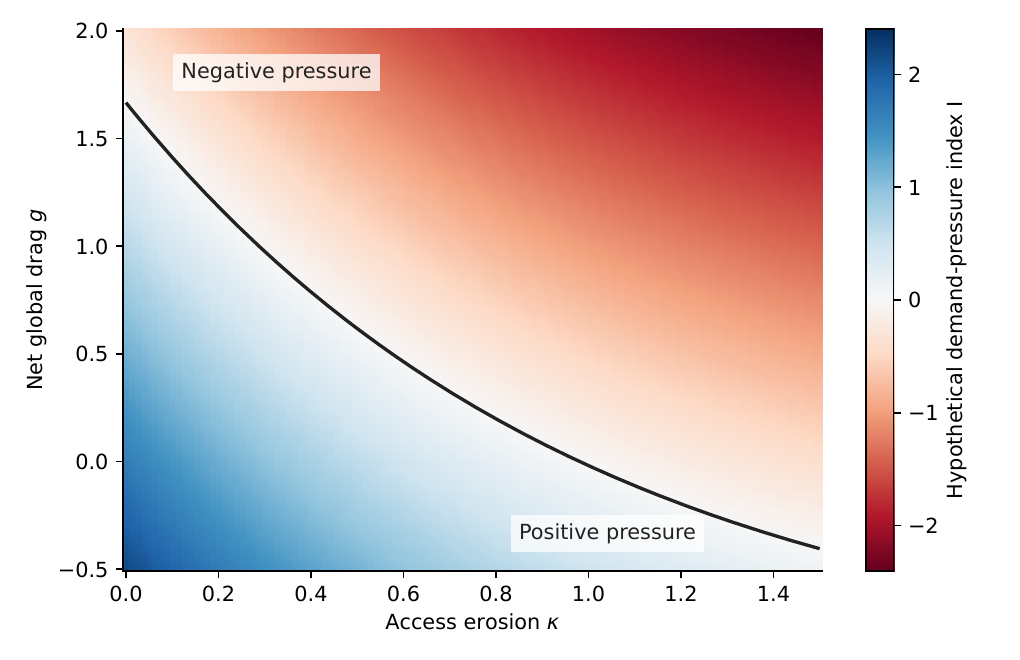}
\caption{Hypothetical sign boundary for $I$ at $z=1$, $p_0=0.2$, $\eta=0.8$, and $\omega=0.4$. The black curve is $I=0$ from \eqref{eq:threshold}. Colors show normalized demand-pressure units, not returns, empirical observations, or probabilities. The grid varies assumed access erosion and net global drag.}
\label{fig:sign}
\end{figure}

With modest access erosion and low global drag, effective BTC demand increases. Keeping the same local demand assumptions but raising global drag reverses the pressure sign. Stronger access erosion can also produce a negative sign despite the same rise in the reference-price BTC share. The comparison demonstrates which restrictions are doing the work; it does not rank the likelihood of the regimes. Figure~\ref{fig:sign} traces the exact finite-shock zero boundary for these fixed remaining parameters.

For the mixed case $(\kappa,g)=(0.6,0.2)$, the index is $0.263422$. Setting $\eta=0$, while holding the other primitives fixed, changes it to $-0.381269$. Setting $\kappa=0$ gives $1.466529$, and setting $g=0$ gives $0.463422$. These channel-removal exercises are model ablations, not causal decompositions of observed returns. The nonlinear changes do not sum to an empirical attribution, and removing a channel may not describe an implementable policy.

For the assumed box $\omega\in[0.2,0.4]$, $\kappa\in[0.2,0.8]$, $\eta\in[0.4,0.8]$, and $g\in[0.1,0.6]$, Proposition~\ref{prop:bounds} gives $I\in[-0.854591,1.083169]$. Both signs remain possible. The box has been chosen to expose that ambiguity; its endpoints have not been inferred from data. Narrower empirically justified restrictions might identify a sign, but numerical precision by itself cannot supply those restrictions.

The accompanying code recreates the two figures, scenario table, and channel-removal values. It checks the analytical log-demand derivative against a local numerical derivative, the exact zero-pressure threshold, and allocation feasibility. These checks establish computational consistency of the illustration, not empirical validity of the assumed elasticities or originality of the framework.

\section{Implications for strategic competition and empirical evaluation}

\subsection{Adjustment regimes and international transmission}

A successful debt resolution can reduce rollover pressure and release bank capacity without canceling all principal. When capital relief is usable and resources are redeployed productively, the reform can support the economic base of strategic competition. When losses instead weaken banks or public services absorb the burden, current strategic effort may remain protected while broader productive capacity deteriorates. It is the allocation of losses and resources, rather than recognized debt alone, that distinguishes these cases.

The allocation block supplies one conditional interpretation of the difference between current strategic spending and future productive capacity. It does not validate any particular strategic prescription. A U.S. response is also endogenous to its own budgets, alliances, technology, and objectives, so Chinese fiscal strain does not mechanically determine relative national power.

The route from this adjustment to Bitcoin is more indirect. Civilian income losses can reduce investable wealth; confidence effects can change desired currency or asset allocation; and access restrictions can prevent execution. At the same time, trade conditions, international risk taking, and U.S. monetary decisions can change the global term in \eqref{eq:price}. A domestic interest-rate reduction or debt swap therefore need not create an equal quantity of internationally accessible dollar funding.

Formally, each resolution regime $r$ selects conditional primitives
\begin{equation}
 \vartheta(r)=\big(c_z(r),\omega(r),\kappa(r),\eta(r),g(r)\big).
 \label{eq:regime}
\end{equation}
The accounting block restricts permissible claims about principal and credit headroom, but it does not identify this vector. A regime that preserves household income can have a different $\omega$ from one that shifts losses to households; a disorderly external response can have a different $g$ even under the same local swap. The numerical regimes compare such conditional mappings rather than claim to derive portfolio elasticities from public-debt accounting.

Dollar stablecoins introduce a further competitive dimension. Greater private demand for dollar-denominated digital balances can extend dollar use rather than substitute for it \cite{bis}. This channel can coexist with interest in nonsovereign Bitcoin, and Proposition~\ref{prop:threshold} shows why the two need not share a price-pressure sign. The framework does not infer that Chinese authorities will adopt Bitcoin as a fiscal solution, that stablecoin growth eliminates capital controls, or that a domestic debt stock can be transferred to token holders without a specified balance-sheet transaction.

Timing matters. A near-term liquidation can produce negative pressure even if a longer-run preference shift would be positive under restored access and improved funding. Conversely, favorable initial liquidity can mask subsequent income losses or tighter restrictions. The model treats each horizon as a conditional state with specified primitives; the numerical $z$ path is not a dated forecast. A richer dynamic model would need to endogenize the policy response, beliefs, market depth, and adjustment lags before assigning a term structure of effects.

\subsection{Observable restrictions and identification limits}

The propositions imply several disciplined empirical comparisons. Under the capital-relief mechanism, otherwise comparable banks with larger eligible exposure and a binding capital constraint should experience more usable headroom when risk-weight relief exceeds losses. Testing that mechanism requires actual regulatory treatment, bank exposure, capital changes, funding, and credit demand. A reduction in reported LGFV debt or a rise in government bond holdings alone does not establish crowding in. The historical evidence in \cite{li} cannot replace these measurements for a later program.

The access mechanism requires distinguishing intentions from executed transactions. An increase in survey demand or an exchange premium can coexist with lower accessible dollar allocations. A test of the preference channel must therefore measure the BTC--dollar-claim composition alongside access, rather than read it from BTC prices alone. Stablecoin turnover is especially inadequate as a proxy for new wealth: the same balance can trade repeatedly, while issuance and redemption can offset across users. Address geography and exchange domicile are also imperfect indicators of beneficial ownership.

For asset prices, a credible study would require unexpected fiscal news, its original release time, synchronized UTC market prices, and an explicit treatment of global confounders. Slow-moving debt levels and a research report's publication date are not automatically exogenous shocks. The November 2024 swap announcement, for instance, followed a November 7 FOMC decision \cite{swap,fed}; a broad event window cannot assign every BTC movement to Chinese debt news. Controls for global equity conditions, dollar funding, macroeconomic surprises, and crypto-specific institutional flows must be conceptually separated from mediators when defining the causal estimand.

These observations define empirical restrictions, not completed tests. A future design might combine consistent local liability panels and pre-event bank exposures for the fiscal block, while using separately identified news innovations for the financial block. Linking the two would require an explicit exclusion or mediation strategy. Regressing a global, continuously traded asset on many provincial debt series does not create independent global price observations, and a provincial difference-in-differences design by itself cannot identify a single global BTC response.

The current paper has no proprietary Wind extract, confidential bank panel, identified Chinese BTC flows, or fitted global-demand equation. It therefore reports no causal coefficient, statistical significance, calibrated expected return, or probability of geopolitical escalation. Its testable content is the set of conditional sign and accounting restrictions. Evidence of a different behavioral or policy response would call for a different specification rather than a retrospective relabeling of the same illustration.

\section{Conclusion}

China's local debt does not map into either strategic weakness or Bitcoin appreciation through a single balance-sheet total. A swap can change recognition and credit capacity without reducing consolidated principal. Protected spending can preserve current strategic effort while concentrating adjustment elsewhere. In the asset block, relative BTC preference, executable demand, dollar stablecoin demand, and global price pressure are different objects. The derived thresholds show exactly how access erosion and global funding conditions can reverse an otherwise favorable preference effect. The framework offers a limited basis for comparing transmission claims and deciding what must be measured. Determining which regime applies, its magnitude, and its probability remains an empirical task.

\section*{Acknowledgments}
OpenAI language models were used to assist in drafting the initial version of this paper.

\begingroup
\raggedright

\endgroup
\end{document}